\documentclass[two column, prA]{revtex4}%
\usepackage{amsmath}
\usepackage{graphicx}
\usepackage{amsfonts}
\usepackage{amssymb}
\usepackage{color}%
\providecommand{\U}[1]{\protect\rule{.1in}{.1in}}
\ifx\pdfoutput\relax\let\pdfoutput=\undefined\fi
\newcount\msipdfoutput
\ifx\pdfoutput\undefined\else
\ifcase\pdfoutput\else
\ifx\paperwidth\undefined\else
\ifdim\paperheight=0pt\relax\else\pdfpageheight\paperheight\fi
\ifdim\paperwidth=0pt\relax\else\pdfpagewidth\paperwidth\fi
\fi\fi\fi
\begin{document}
\title[Photon position observable]{Photon quantum mechanics with a position observable}
\author{Margaret Hawton}
\affiliation{Lakehead University, Thunder Bay, ON, Canada, P7B 5E1}
\email{mhawton@lakeheadu.ca}

\begin{abstract}
We second quantize an explicitly Lorentz invariant lagrangian density and
derive a theory of photon quantum mechanics. The one photon Hilbert space is
the vector space of normalizable positive frequency four-potentials
$\left\vert A^{+}\right\rangle $ with scalar product $\left(  e/\hbar\right)
\left\langle A_{1}^{+}\left\vert \Omega\right\vert A_{2}^{+}\right\rangle $
where $\Omega$ is the nonlocal frequency operator. Observables are described
by the Poincare operators augmented with a photon position operator. It is
found that the probability amplitude to observe a photon in a bounded region
of space, defined as the projection of $\left\vert A^{+}\right\rangle $ onto
the basis of position eigenvectors, equals the inverse Fourier transform of
the probability amplitude for a plane wave. A continuity equation that
describes photon propagation in free space and an optical circuit is derived.

\end{abstract}
\maketitle

\section{Introduction}

In quantum field theory (QFT) photons are discrete excitations of classical
electromagnetic (EM) fields created and annihilated by second quantized
operators \cite{Dirac}. When a stream of single photon pulses is split with a
Fresnel biprism coincidence counts registered by photon counting detectors in
the two arms is strongly reduced \cite{BeamSplitter}. This verifies
experimentally the quantum mechanical nature of single photons. Physical one
photon pulses coupled to optical circuits containing transmission lines, beam
splitters and photon counting detectors are now routinely prepared in the
laboratory \cite{Manufacturable}. This motivated derivation of a photon
continuity equation in \cite{Conserved}.

Absorption in a small photon counting detector is essentially a measurement of
photon position, but there is no general agreement as to how the probability
to count a photon should be calculated. Proposals in the published literature
are energy density \cite{SmithRaymer} or probability density
\cite{Nonlocality} based but these approaches are not compatible since the
energy density of all single photon states is nonlocal
\cite{Nonlocality,DeBievre}. Spohn describes a theory of photon quantum
mechanics (QM) in transverse Hilbert space with conserved observables
represented by the Poincare operators \cite{Spohn} but does not introduce a
position operator. It appears that there is no expression for the probability
to observe a single photon in a bounded region of space in the published literature.

Here the nonlocal density described in \cite{DeBievre,Conserved} is used to
define the scalar product in the vector space of positive frequency
four-potentials while position is identified as an obervable in this Hilbert
space. The probability amplitide to observe a particular photon is identified
as the projection of its state vector onto a basis of position eigenvectors
\cite{HawtonDebierre}. These positon eigenvectors are cylindrically
symmetrical consistent with the geometry of the photon Wigner little group.
They are eigenvectors of a position operator that does not transform as a
simple vector because a term must be added that rotates their axis of symmetry.

In the next Section we will second quantize an explicitly Lorentz invariant
Lagrangian density, impose the Lorenz gauge condition and derive a scalar
product for the $\left\vert A^{+}\right\rangle $ Hilbert space. We then
calculate the Poincare and position eigenvectors and interpret the projection
of the state vector onto the position eigenvectors as probability amplitudes
to observe a photon. Based on the covariant equation of motion, we derive a
continuity equation for photon propogation in free space and an optical
circuit with local coupling to electrical currents in the source and detector.

\section{Theory}

Based on the homogeneous Maxwell equations (MEs)
\begin{equation}
\nabla\cdot\mathbf{B}\left(  x\right)  =0,\ \nabla\times\mathbf{E}\left(
x\right)  +\partial_{t}\mathbf{B}\left(  x\right)  =0 \label{ME12}%
\end{equation}
an EM four-potential $A=\left(  \phi/c,\mathbf{A}\right)  $ can be defined
such that
\begin{align}
\mathbf{B}\left(  x\right)   &  =\nabla\times\mathbf{A}\left(  x\right)
,\label{B}\\
\mathbf{E}\left(  x\right)   &  =-\partial_{t}\mathbf{A}\left(  x\right)
-\nabla\phi\left(  x\right)  \label{E}%
\end{align}
and%
\begin{equation}
F^{\mu\nu}\left(  x\right)  =\partial^{\mu}A^{\nu}-\partial^{\nu}A^{\mu}
\label{Faraday}%
\end{equation}
is the Faraday tensor. SI units are used throughout. The contravariant
space-time, four-potential, wavevector, momentum and electric current
four-vectors are $x=x^{\mu}=\left(  ct,\mathbf{x}\right)  ,$ $A=\left(
\phi/c,\mathbf{A}\right)  ,$ $k=\left(  \omega_{k}/c,\mathbf{k}\right)  $,
$p=\hbar k$ and $J_{e}=\left(  c\rho_{e},\mathbf{j}_{e}\right)  $, the
four-gradiant is $\partial^{\mu}=\left(  \partial_{ct},-\mathbf{\nabla
}\right)  $, $\square\equiv\partial_{ct}^{2}-\mathbf{\nabla}^{2}$,
$h=2\pi\hbar$ is Planck's constant and $c$ is the speed of light in free
space. The covariant four-vector corresponding to $U^{\mu}$ is $U_{\mu}%
=g_{\mu\nu}U^{\nu}=\left(  U_{0},-\mathbf{U}\right)  $ where $g$ is a
$4\times4$ diagonal matrix with diagonal $\left(  1,-1,-1,-1\right)  $.

We base our theory on the Lagrangian density%
\begin{equation}
\mathcal{L}_{cov}=-\frac{1}{2}\left(  \partial_{\mu}A_{\nu}\right)  \left(
\partial^{\mu}A^{\nu}\right)  -J_{e\mu}A^{\mu}\label{cov}%
\end{equation}
in the Lorenz gauge where%
\begin{equation}
\Lambda\equiv\partial_{\mu}A^{\mu}=\frac{1}{c^{2}}\partial_{t}\phi
+\mathbf{\nabla}\cdot\mathbf{A}=0.\label{lorenz}%
\end{equation}
$\mathcal{L}_{cov}$ is equivalent to the covariant gauge fixed Maxwell
Lagrangian density $\mathcal{L=-\varepsilon}_{0}c^{2}\left(  \frac{1}{4}%
F_{\mu\nu}F^{\mu\nu}+\frac{1}{2}\Lambda^{2}\right)  -J_{e\mu}A^{\mu}$
\cite{CT}. The equation of motion derived from (\ref{cov}) is%
\begin{equation}
\varepsilon_{0}\square A^{\nu}=J_{e}^{\nu}.\label{motion}%
\end{equation}
Since $\Lambda=0$ (\ref{motion}) is consistent with the MEs
\begin{align}
\square\phi-\partial_{t}\Lambda &  =\mathbf{\nabla}\cdot\mathbf{E}=\rho
_{e}/\varepsilon_{0},\label{ME3}\\
\square\mathbf{A}+\mathbf{\nabla}\Lambda &  =-\mu_{o}\varepsilon_{0}%
\partial_{t}\mathbf{E}+\mathbf{\nabla}\times\mathbf{B=}\mu_{0}\mathbf{j}%
_{e}.\label{ME4}%
\end{align}
The momentum conjugate to $A^{\mu}$ is
\begin{align}
\Pi^{\mu} &  =\partial\mathcal{L}_{cov}/\partial\left(  \partial_{t}A_{\mu
}\right)  \nonumber\\
&  =\varepsilon_{0}\partial_{t}A^{\mu}.\label{conjugate}%
\end{align}
Based on the usual rules of second quantization the norm of a one-photon state
is $\left[  \widehat{A}_{\mu},\widehat{\Pi}^{\mu}\right]  /i\hbar=1$.

QM is formulated in a Hilbert space with an inner product that selects a time
axis \cite{Zajac} and hence a particular Lorentz reference frame. The one
photon Hilbert space will be defined as the vector space of normalizable field
vectors $\left\vert A^{+}\right\rangle $ that satisfy the subsiduary condition
$\partial_{\mu}A^{\mu+}\left\vert 0\right\rangle =0$ \cite{CT}. The scalar
product is
\begin{align}
\left\langle A_{1}^{+}\left\vert \varepsilon_{0}\Omega/\hbar\right\vert
A_{2}^{+}\right\rangle  &  =\frac{i\varepsilon_{0}}{2\hbar}\int_{t}%
d\mathbf{x}A_{1\mu}\partial_{t}A_{2}^{\mu}+H.c.\label{sp_mu}\\
&  =\frac{\varepsilon_{0}}{\hbar}\left\langle \psi_{LP1}|\psi_{LP2}%
\right\rangle \text{.}\label{sp_LP}%
\end{align}
where the Hermitian nonlocal frequency operator is \cite{DeBievre}
\begin{align}
\Omega & =c\left(  -\mathbf{\nabla}^{2}\right)  ^{1/2},\label{Omega}\\
i\hbar\partial_{t}\left\vert \psi\right\rangle  & =\widehat{H}\left\vert
\psi\right\rangle =\hbar\Omega\left\vert \psi\right\rangle \label{SE}%
\end{align}
is the Schrodinger time evolution equation and the Landau-Peierls wave
function is \cite{LP,Nonlocality}%
\begin{equation}
\text{ }\psi_{LP}=\sqrt{\frac{\varepsilon_{0}}{2\hbar}}\left[  \Omega
^{1/2}\mathbf{A}-i\Omega^{-1/2}\mathbf{E}\right]  =\sqrt{\frac{2\varepsilon
_{0}}{\hbar}}\Omega^{1/2}A^{+}.\label{LP}%
\end{equation}
The photon bosonic Fock space is the direct sum of the symmetric tensor
products of all $n$-photon Hilbert spaces built from this single photon
Hilbert space. For a one photon state we set $\left\langle A^{+}\left\vert
\varepsilon_{0}\Omega/\hbar\right\vert A^{+}\right\rangle =1$. The integrand
in (\ref{sp_mu}) is proportional to the time-like component of the covariant
photon four-current.

Plane waves are eigenvectors of the momentum and energy operators. In position
space the momentum and energy eigenvector equations are%
\begin{align}
\widehat{\mathbf{p}}e^{-ikx} &  =-i\hbar\mathbf{\nabla}e^{-ikx}=\hbar
\mathbf{k}e^{-ikx}\label{peq}\\
\widehat{H}e^{-ikx} &  =\hbar\Omega e^{-ikx}=\hbar kce^{-ikx}\label{Heq}%
\end{align}
where $-kx=\mathbf{k\cdot x}-\omega_{k}t$ with $\omega_{k}=kc$ in free space.
The Hamilton operator $\widehat{H}=\hbar\Omega$ is nonlocal. As in Schrodinger
QM plane waves are a useful basis but they are not normalizable so they are
not physical states. The covariant positive frequency four-potential can be
written as \
\begin{equation}
A_{\mu}^{+}\left(  x\right)  =i\sqrt{\frac{\hbar}{2\varepsilon_{0}}}\int%
\frac{d\mathbf{k}}{\left(  2\pi\right)  ^{3/2}\omega_{k}^{1/2}}c_{\mu}\left(
\mathbf{k}\right)  e^{-ikx}\label{4potential}%
\end{equation}
where $c_{\mu}\left(  \mathbf{k}\right)  $ is the probability amplitude for
wave vector $\mathbf{k}$. In momentum space the scalar product is%
\begin{equation}
\left\langle A_{1}^{+}\left\vert \varepsilon_{0}\Omega/\hbar\right\vert
A_{2}^{+}\right\rangle =\int_{t}\frac{d\mathbf{k}}{\left(  2\pi\right)  ^{3}%
}c_{1\mu}^{\ast}\left(  \mathbf{k}\right)  c_{2}^{\mu}\left(  \mathbf{k}%
\right)  .\label{sp_mu_k}%
\end{equation}

The mutually orthogonal unit vectors are
\begin{align}
e_{0} &  =\left(  1,0,0,0\right)  ,\nonumber\\
\mathbf{e}_{3} &  =\mathbf{e}_{\mathbf{k}}=\mathbf{k}/\left\vert
\mathbf{k}\right\vert \nonumber\\
\mathbf{e}_{\lambda}\left(  \mathbf{k}\right)   &  =\frac{1}{\sqrt{2}}\left(
\mathbf{e}_{\theta}+i\lambda\mathbf{e}_{\phi}\right)  \text{ for }\lambda
=\pm1.\label{basis}%
\end{align}
Since $c_{0}\left(  \mathbf{k}\right)  =c_{3}\left(  \mathbf{k}\right)  $ in
the Lorenz gauge \cite{IZ,CT,GuptaBleuler} these terms cancel in
(\ref{sp_mu_k}) to give%
\begin{align}
\left\langle A_{1}^{+}\left\vert \varepsilon_{0}\Omega/\hbar\right\vert
A_{2}^{+}\right\rangle  &  =\sum_{\lambda=\pm1}\int_{t}\frac{d\mathbf{k}%
}{\left(  2\pi\right)  ^{3}}c_{1\lambda}^{\ast}\left(  \mathbf{k}\right)
c_{2\lambda}\left(  \mathbf{k}\right)  \label{spk}\\
&  =\frac{i\varepsilon_{0}}{2\hbar}\int_{t}d\mathbf{xA}_{1\perp}\cdot
\partial_{t}\mathbf{A}_{2\perp}+H.c..\label{spx}%
\end{align}
This is consistent with the $\mathbf{k}$-space scalar product derived from the
standard Lagrangian in the Coulomb gauge in Spohn \cite{Spohn}. The state
(\ref{4potential}) is physically equivalent to the transverse potential
\cite{Morrissey}%
\begin{equation}
\mathbf{A}^{+}\left(  x\right)  =i\sqrt{\frac{\hbar}{2\varepsilon_{0}}}%
\sum_{\lambda=\pm1}\int\frac{d\mathbf{k}}{\left(  2\pi\right)  ^{3/2}%
\omega_{k}^{1/2}}\mathbf{e}_{\lambda}\left(  \mathbf{k}\right)  c_{\lambda
}\left(  \mathbf{k}\right)  e^{-ikx}.\label{Aperp}%
\end{equation}

The choice $c_{1\lambda}\left(  \mathbf{k}\right)  =e^{i\left(  \mathbf{k}%
\cdot\mathbf{x}^{\prime}-kct\right)  }$ and $c_{2\lambda^{\prime}}\left(
\mathbf{k}\right)  =e^{i\left(  \mathbf{k}\cdot\mathbf{x}-kct\right)  }$ for
$\lambda,\lambda^{\prime}=\pm1$ substituted into (\ref{spk}) to give%
\begin{equation}
\left\langle A_{\mathbf{x\lambda}}^{+}\left\vert \varepsilon_{0}\Omega
/\hbar\right\vert A_{\mathbf{x}^{\prime}\lambda^{\prime}}^{+}\right\rangle
=\delta_{\lambda\lambda^{\prime}}\delta(\mathbf{x}-\mathbf{x}^{\prime
})\label{xcomm}%
\end{equation}
describes a basis of exactly localized states in the helicity basis. In
momentum space the position operator is \cite{HawtonDebierre}%
\begin{equation}
\widehat{\mathbf{x}}^{\left(  \alpha\right)  }=i\mathbf{\nabla}_{\mathbf{k}%
}-i\alpha\mathbf{k}/k^{2}+\mathbf{k}\times\mathbf{S}/k+cot\left(
\theta\right)  \mathbf{e}_{\phi}/k.\label{xop}%
\end{equation}
The eigenvectors of $\widehat{\mathbf{x}}^{\left(  -1/2\right)  },$
$\widehat{\mathbf{x}}^{\left(  1/2\right)  }=\widehat{\mathbf{x}}^{\left(
-1/2\right)  \dag}$ and $\widehat{\mathbf{x}}^{\left(  0\right)
}=\widehat{\mathbf{x}}^{\left(  0\right)  \dag}$ are $\left\vert
A^{+}\right\rangle $, $\left\vert \partial_{t}A^{+}\right\rangle $ and
$\left\vert \psi_{LP}\right\rangle $ respectively. If the scalar product is
evaluated using (\ref{sp_LP}) the position operator is Hermitian. The position
eigenvectors are cylindrically symmetric with helicity and total angular
momentum parallel to their direction of propagation as required by the
symmetry of the photon little group \cite{Weinberg}. \ For $3$-axis chosen
parallel to the direction of photon propagation the position operators satisfy
the commutation relations $\left[  \widehat{J}_{3},\widehat{x}_{1}\right]
=i\widehat{x}_{2}$, $\left[  \widehat{J}_{3},\widehat{x}_{2}\right]
=-i\widehat{x}_{1}$ and $\left[  \widehat{x}_{1},\widehat{x}_{2}\right]  =0$
\cite{HawtonDebierre} and are well suited to the description of "twisted
light" \cite{Twisted}. The probability amplitude for position $\mathbf{x}$,%
\begin{equation}
\psi\left(  t,\mathbf{x}\right)  \equiv\left\langle A_{\mathbf{x}\mu}%
^{+}|A^{+}\right\rangle =\int_{t}\frac{d\mathbf{k}}{\left(  2\pi\right)  ^{3}%
}c_{\mu}\left(  \mathbf{k}\right)  e^{-i\mathbf{k\cdot x}},\label{xampl}%
\end{equation}
is the inverse Fourier transform of $c_{\mu}\left(  \mathbf{k}\right)  $. The
probability density to observe a photon at position $\mathbf{x}$ is
$\left\vert \psi\left(  t,\mathbf{x}\right)  \right\vert ^{2}$ and $\int%
_{R}d\mathbf{x}\left\vert \psi\left(  t,\mathbf{x}\right)  \right\vert ^{2}$
is the probability to observe the photon in the bounded region of space $R$ at
time $t$..

The plane wave creation and annihilation operators provide a useful
description of photon Fock space. The covariant annihilation and creation
operators satisfy
\begin{align}
\widehat{a}_{\mathbf{k},0}\left\vert 0\right\rangle  &  =\widehat{a}%
_{\mathbf{k},3}\left\vert 0\right\rangle \label{Lorenz}\\
\left[  \widehat{a}_{\mathbf{k}\mu},\widehat{a}_{\mathbf{k}^{\prime}\nu
}\right]   &  =0,\ \left[  \widehat{a}_{\mathbf{k}\mu}^{\dag},\widehat{a}%
_{\mathbf{k}^{\prime}\nu}^{\dagger}\right]  =0\label{commute}\\
\left[  \widehat{a}_{\mathbf{k}\mu},\widehat{a}_{\mathbf{k}^{\prime}\nu
}^{\dagger}\right]   &  =-g_{\mu\nu}\delta_{\mathbf{kk}^{\prime}%
}\label{dontcommute}\\
\ \left\vert n_{\mathbf{k}\mu}\right\rangle  &  =\left(  \widehat{a}%
_{\mathbf{k}\mu}\right)  ^{n}\left\vert 0\right\rangle ,\label{nphotonstate}\\
\left\langle n_{\mathbf{k}\mu}\left\vert \widehat{a}_{\mathbf{k}\mu}^{\dag
}\widehat{a}_{\mathbf{k}\mu}\right\vert n_{\mathbf{k}\mu}\right\rangle  &
=n_{\mathbf{k}\mu},\ \\
\left\langle n_{\mathbf{k}\mu}\left\vert \widehat{a}_{\mathbf{k}\mu
}\widehat{a}_{\mathbf{k}\mu}^{\dag}\right\vert n_{\mathbf{k}\mu}\right\rangle
&  =\left\langle n_{\mathbf{k}\mu}+1|n_{\mathbf{k}\mu}+1\right\rangle
=n_{\mathbf{k}\mu}-g_{\mu\mu}%
\end{align}%
\begin{equation}
\left\langle n_{\mathbf{k}\mu}\left\vert \widehat{a}_{\mathbf{k}\mu
}\widehat{a}_{\mathbf{k}^{\prime}\nu}^{\dagger}-\widehat{a}_{\mathbf{k}%
^{\prime}\nu}^{\dagger}\widehat{a}_{\mathbf{k}\mu}\right\vert n_{\mathbf{k}%
\mu}\right\rangle =-g_{\mu\nu}.\label{countsone}%
\end{equation}
Eq. (\ref{countsone}) implies that the commutator does not depend on
$\left\vert n_{\mathbf{k}\mu}\right\rangle $ so (\ref{dontcommute}) describes
the addition of one photon to Fock state. In the continuum limit $\Delta
n/V\rightarrow d\mathbf{k}/\left(  2\pi\right)  ^{3}$ so we define the plane
wave basis
\begin{equation}
\left[  \widehat{a}_{\mu}\left(  \mathbf{k}\right)  ,\widehat{a}_{\nu
}^{\dagger}\left(  \mathbf{k}^{\prime}\right)  \right]  =-g_{\mu\nu}%
\delta\left(  \mathbf{k-k}^{\prime}\right)  \label{kcommutation}%
\end{equation}
and the four-potential operator \
\begin{align}
\widehat{A}_{\mu}^{+}\left(  x\right)   &  =i\sqrt{\frac{\hbar}{2\varepsilon
_{0}}}\int\frac{d\mathbf{k}}{\left(  2\pi\right)  ^{3/2}\omega_{k}^{1/2}%
}\nonumber\\
&  \times\widehat{a}_{\mu}\left(  \mathbf{k}\right)  c_{\mu}\left(
\mathbf{k}\right)  e^{-ikx}.\label{Aplusop}%
\end{align}
The superscript $\dagger$ is the Hermitian conjugate, $\pm$ refer to positive
and negative frequency parts and $c_{\mu}\left(  \mathbf{k}\right)  $ is the
probability amplitude for wave vector $\mathbf{k}$. In strictly covariant form
the commutators and $\mathbf{k}$-space volume should be replaced with $\left[
\widehat{a}_{\mu}\left(  \mathbf{k}\right)  ,\widehat{a}_{\nu}^{\dagger
}\left(  \mathbf{k}^{\prime}\right)  \right]  \rightarrow-g_{\mu\nu}%
\omega^{1/2}\left(  2\pi\right)  ^{3/2}\delta\left(  \mathbf{k-k}^{\prime
}\right)  $ and $d\mathbf{k/}\left(  2\pi\right)  ^{3}\omega_{k}%
^{1/2}\mathbf{\rightarrow}d\mathbf{k}/\mathbf{\left(  2\pi\right)  }^{3}%
2k^{0}$ \cite{IZ}.

The covariant second quantized equations of motion are
\begin{equation}
\varepsilon_{0}\square\widehat{A}^{\mu}=\widehat{J}_{e}^{\mu}%
.\label{2ndquantEqsofMotioh}%
\end{equation}
As in \cite{Conserved} a continuity equation for the conserved photon current
operator can be derived from commutator of $\widehat{A}$ with the equation of
motion, here (\ref{2ndquantEqsofMotioh}), to give%
\begin{equation}
\left[  \widehat{A}_{\mu},\varepsilon_{0}\square\widehat{A}^{\mu}\right]
=\varepsilon_{0}\partial_{\nu}\left[  \widehat{A}_{\mu},\partial^{\nu
}\widehat{A}^{\mu}\right]  =\left[  \widehat{A}_{\mu},\widehat{J}_{e}^{\mu
}\right]  ,\label{FermiCommutation}%
\end{equation}
and the continuity equation%
\begin{equation}
\frac{-i\varepsilon_{0}}{2\hbar}\left(  \partial_{t}\left[  \widehat{A}_{\mu
},\partial_{t}\widehat{A}^{\mu}\right]  +\mathbf{\nabla\cdot}\left[
\widehat{A}_{\mu},\mathbf{\nabla}\widehat{A}^{\mu}\right]  \right)
=\frac{-i\varepsilon_{0}}{2\hbar}\left[  \widehat{A}_{\mu},\widehat{J}%
_{e}^{\mu}\right]  .\label{Continuity}%
\end{equation}
The photon four-current operator derived from the Lagrangian (\ref{cov}) is
then%
\begin{equation}
\widehat{J}_{cov}^{\mu}=\left(  c\widehat{\rho}_{p},\widehat{\mathbf{J}}%
_{p}\right)  =\frac{-i\varepsilon_{0}}{2\hbar}\left(  \left[  \widehat{A}%
_{\mu},\partial_{t}\widehat{A}^{\mu}\right]  ,\left[  \widehat{A}_{\mu
},\mathbf{\nabla}\widehat{A}^{\mu}\right]  \right)  \label{Jopcov}%
\end{equation}
In Gupta-Bleuler quantization the Lorenz subsidiary condition forces scalar
and longutudinal photon excitations to appear in matched numbers so these
terms cancel in $\widehat{J}_{cov}^{\mu}$ leaving only the  transverse modes.
In the helicity basis (\ref{Continuity}) reduces to%
\begin{align}
&  \frac{-i\varepsilon_{0}}{2\hbar}\sum_{\lambda=\pm1}\left\{  \partial
_{t}\left(  \widehat{\mathbf{A}}_{\lambda}\cdot\partial_{t}\widehat{\mathbf{A}%
}_{\lambda}\right)  +\mathbf{\nabla\cdot}\left(  \widehat{A}_{\lambda
},\mathbf{\nabla}\widehat{A}_{\lambda}\right)  \right\}  +H.c.\nonumber\\
&  =\frac{-i\varepsilon_{0}}{2\hbar}\left(  \widehat{\mathbf{A}}%
\cdot\widehat{\mathbf{J}}_{e}-\widehat{\phi}\widehat{\rho}_{e}\right)
+H.c..\label{lambdaContinuity}%
\end{align}
where%
\begin{equation}
\widehat{\mathbf{E}}_{\lambda}=-\partial_{t}\widehat{\mathbf{A}}_{\lambda
}.\label{Elambda}%
\end{equation}

In the continuity equation (\ref{Continuity}) polarization and magnetization
of the medium act as external driving forces, however many recent experiments
involve propagation in transmission lines and optical circuits. At infrared
and visible frequencies a medium can be treated as continuous by averaging
over domains of order $10^{-8}m$ \cite{Conserved}. In the presence of
localized sources and sinks described by $\widehat{J}_{es}$ the electric
current operator can be written as%
\begin{align}
\mathbf{J}_{e} &  =\partial_{t}\mathbf{P}+\mathbf{\nabla}\times\mathbf{M}%
+\mathbf{J}_{es},\label{PandM}\\
\widehat{J}_{es} &  =\left(  c\widehat{\rho}_{es},\ \widehat{\mathbf{J}}%
_{es}\right)  .\label{barerho}%
\end{align}
where $\mathbf{P}$ is polarization, $\mathbf{M}$ is magnetization and electric
displacement and magnetic field operators are
\begin{align}
\mathbf{D} &  =\varepsilon_{0}\mathbf{E}+\mathbf{P}=\varepsilon\mathbf{E}%
,\label{D}\\
\mathbf{H} &  =\mu_{0}^{-1}\mathbf{B}-\mathbf{M}=\mu^{-1}\mathbf{B.}\nonumber
\end{align}
Since $\mathbf{\nabla}\cdot\mathbf{D}=0$ in a charge free region
(\ref{lambdaContinuity}) can be written as%
\begin{equation}
\partial_{t}\widehat{\rho}_{pm}\left(  x\right)  \mathbf{+\nabla}%
\cdot\widehat{\mathbf{J}}_{pm}\left(  x\right)  =\frac{-i\varepsilon_{0}%
c}{2\hbar}\widehat{A}_{\mu}\widehat{J}_{es}^{\mu}+H.c.\label{Jmcontinuity}%
\end{equation}
where%
\begin{equation}
\widehat{J}_{pm}\left(  x\right)  =\frac{-i}{2\hbar}\left(
\widehat{\mathbf{A}}\cdot\widehat{\mathbf{D}},-c\varepsilon\mu
\widehat{\mathbf{A}}\times\widehat{\mathbf{H}}\right)  +H.c.\label{Jm}%
\end{equation}
is the four-current operator in a medium describing propagation at speed
$v=c/n=1/\sqrt{\varepsilon\mu}$. This is the general case since it describes
propagation in free space if $\varepsilon=\varepsilon_{0}$ and $\mu=\mu_{0}$.

In (\ref{Jmcontinuity}) $\widehat{J}_{es}^{\mu}$ is the electric four-current
operator describing the source or detector and $\widehat{A}_{\mu}%
^{-}\widehat{J}_{es}^{\mu+}+H.c.$ describes Rabi oscillations in the rotating
wave approximation. In the source an atom or quantum dot is prepared in an
excited state so that it will emit a transverse photon. In a detector the
semiconducting device is prepared in its ground state and when a photon is
absorbed an electric field separates the electron-hole pair, amplifies the
current in an external circuit and the detector "clicks". If the detector is
small this is essentially a measurement of photon position. A photon is
annihilated so that the number of photons in Fock space is reduced by $1$. A
scattering experiment can be analyzed in the momentum basis. When a photon
pulse containing a wide range of frequencies with momentum space probability
amplitudes $c_{\lambda}\left(  \mathbf{k}\right)  $ is scattered each
frequency scatters independently.

There are several good reasons to doubt that a photon is localizable. Any
positive frequency function initially localized in a finite region spreads
instantaneously throughout space \cite{Hegerfeldt} and there are no local
annihilation or creation operators \cite{RS}. A coordinate space scalar
product defined in terms of energy density is nonlocal \cite{SmithRaymer}. The
absolute square of the Bialynicki-Birula-Sipe photon wave function,
$\mathbf{F}=\sqrt{\varepsilon_{0}/2}\left(  \mathbf{E}+i\lambda c\mathbf{B}%
\right)  $, is energy density. Federico and Jauslin \cite{FedericoJauslin}
define a general single-photon Landau-Peierls (LP) field $\psi=\sqrt
{\frac{\varepsilon_{0}}{2\hbar}}\left(  \Omega^{1/2}\mathbf{A}-i\Omega
^{-1/2}\mathbf{E}\right)  $. Since $\mathbf{F}=i\sqrt{\hbar}\Omega
^{1/2}\mathbf{\psi}_{LP}$ they concluded that the energy density of any single
photon state is nonlocal. The sign of frequency operator $\widehat{\epsilon
}=i\left(  \Omega/c\right)  \partial_{ct}$ was defined in
\cite{Mostafazadeh,Validation} to make $\left(  \varepsilon_{0}/\hbar\right)
\left(  \widehat{\epsilon}\mathbf{E}\right)  \cdot\mathbf{A}$ positive
definite, but this operator is also nonlocal. Initially single photon pulses
were modelled classically using real fields \cite{LocalPhotons}.

Here the integral over all space of the nonlocal density is used to define the
scalar product and norm of a one photon state. The Hamiltonian operator that
describes time evolution is nonlocal, but propagation at the speed of light is
described by a continuity equation consistent with MEs.  The photon
probability amplitude is the projection of its state vector onto a basis of
positon eigenvectors which equals the inverse Fourier transform of the plane
wave probability amplitude. Its absolute square can be integrated over a
bounded region to give the probability to count a photon and answer Debievre's
question "where's that quantum" \cite{DeBievre}.

\section{Conclusion}

Second quantization of an explicitly invariant Lagrangian in the Lorenz gauge
leads to a one photon Hilbert space of positive frequency four-potentials
$\left\vert A^{+}\right\rangle $ with scalar product $\left\langle A_{1}%
^{+}\left\vert \varepsilon_{0}\Omega/\hbar\right\vert A_{2}^{+}\right\rangle $
that can be extended to a Fock space that is the direct sum of the symmetric
tensor products of all $n$-photon Hilbert spaces. The Poincare operators
augmented with a photon position operator with commuting components represent
quantum mechanical observables. Absorption of a photon in a small photon
counting detector is essentially a measurement of positon and it was proved
here that the probability to count a photon is the integral of the absolute
square of the inverse Fourier transform of the plane wave probability
amplitude over the volume of the detector.

\end{document}